\documentclass[aps,prl,twocolumn,preprintnumbers,superscriptaddress]{revtex4-2}

\pdfoutput=1

\usepackage[T1]{fontenc} 
\usepackage{graphicx}
\usepackage{epsfig}
\usepackage{amssymb}
\usepackage{amsfonts}
\usepackage{amsmath,euscript,array,mathrsfs}
\usepackage{bbold}
\usepackage{epsf}
\usepackage{slashed}
\usepackage{comment}
\usepackage{colortbl}

\usepackage[caption=false]{subfig}
\usepackage[colorlinks=true,linktocpage=true,linkcolor=blue,citecolor=blue]{hyperref}

\newcommand{\be}{\begin{equation}}
\newcommand{\ee}{\end{equation}}
\newcommand{\beqs}{\begin{eqnarray}}
\newcommand{\eeqs}{\end{eqnarray}}

\begin{document}


\preprint{ICCUB-21-004}

\title{Bubble Wall Velocity from Holography}

\author{Yago Bea}
\affiliation{School of Mathematical Sciences, Queen Mary University of London, Mile End Road, London E1 4NS, United Kingdom}
\author{Jorge Casalderrey-Solana}
\affiliation{Departament de F\'\i sica Qu\`antica i Astrof\'\i sica \&  Institut de Ci\`encies del Cosmos (ICC), Universitat de Barcelona, Mart\'{\i}  i Franqu\`es 1, 08028 Barcelona, Spain}
\author{Thanasis Giannakopoulos}
\affiliation{Centro de Astrof\'{\i}sica $E$ Gravita\c c\~ao -- CENTRA, Departamento de F\'{\i}sica, Instituto Superior T\'ecnico -- IST, Universidade de Lisboa -- UL, Av.\ Rovisco Pais 1, 1049-001 Lisboa, Portugal}
\author{David Mateos}
\affiliation{Departament de F\'\i sica Qu\`antica i Astrof\'\i sica \&  Institut de Ci\`encies del Cosmos (ICC), Universitat de Barcelona, Mart\'{\i}  i Franqu\`es 1, 08028 Barcelona, Spain}
\affiliation{Instituci\'o Catalana de Recerca i Estudis Avan\c cats (ICREA), 
Llu\'\i s Companys 23, Barcelona, Spain}
\author{Mikel Sanchez-Garitaonandia}
\affiliation{Departament de F\'\i sica Qu\`antica i Astrof\'\i sica \&  Institut de Ci\`encies del Cosmos (ICC), Universitat de Barcelona, Mart\'{\i}  i Franqu\`es 1, 08028 Barcelona, Spain}
\author{Miguel Zilh\~ao}
\affiliation{Centro de Astrof\'{\i}sica $E$ Gravita\c c\~ao -- CENTRA,
  Departamento de F\'{\i}sica, Instituto Superior T\'ecnico -- IST, Universidade
  de Lisboa -- UL, Av.\ Rovisco Pais 1, 1049-001 Lisboa, Portugal}


\begin{abstract}
Cosmological phase transitions proceed via the nucleation of bubbles that subsequently expand and collide. The resulting gravitational wave spectrum depends crucially on the bubble wall velocity. We use holography to compute the wall velocity from first principles in a strongly coupled, non-Abelian, four-dimensional gauge theory. The wall velocity is determined dynamically in terms of the nucleation temperature. We verify that ideal hydrodynamics provides a good description of the system everywhere except near the wall.
\end{abstract}


\maketitle

\noindent
{\bf Introduction.}
Although the Electroweak (EW) transition is believed to be a smooth crossover \cite{Kajantie:1996mn,Laine:1998vn,Rummukainen:1998as} in the Standard Model (SM), this turns into a first-order phase transition (PT) even in minimal extensions thereof \cite{Carena:1996wj,Delepine:1996vn,Laine:1998qk,Huber:2000mg,Grojean:2004xa,Huber:2006ma,Profumo:2007wc,Barger:2007im,Laine:2012jy,Dorsch:2013wja,Damgaard:2015con}. In these scenarios the Universe  undergoes this PT as it expands and cools, and this results in the production of gravitational waves (GWs) \footnote{For a review see e.g.~\cite{Hindmarsh:2020hop}.} potentially observable by detectors such as LISA \cite{Caprini:2019egz}. Further scenarios that motivate the study of cosmological PTs include Grand Unified Theories \cite{Georgi:1974sy,Pati:1974yy} and strongly interacting Dark Matter (DM) \cite{Kribs:2016cew,Tulin:2017ara}. In the first case one imagines a gauge theory defined at an energy scale much higher than the EW scale that could have its own PTs \cite{Guth:1981uk,Kuzmin:1982uq,Huang:2020bbe}. The second case corresponds to the possibility that DM, while weakly interacting with the SM, might be strongly interacting with itself. Large classes of models in this category possess first-order PTs and lead to the production of GWs \cite{Schwaller:2015tja,Huang:2020mso}. In summary, the discovery of GWs originating from a cosmological PT would not only be the discovery of new physics beyond the SM but, in some cases, it may be our only realistic window into such physics. 

Maximising the discovery potential  requires an accurate prediction of the GW spectrum. The PT proceeds via the nucleation of bubbles of the stable, low-energy phase inside the supercooled phase. These bubbles subsequently expand and collide. The computation of the resulting GW spectrum requires knowledge of several parameters. Some of these, such as the critical temperature, the strength of the transition, etc are thermodynamic in nature and can be computed from static properties of the underlying theory \footnote{Holographic calculations of some of these parameters include \cite{Ahmadvand:2017xrw,Attems:2017ezz,Ahmadvand:2017tue,Bea:2018whf,Attems:2018gou,Attems:2019yqn,Ahmadvand:2020fqv,Bea:2020ees,Bigazzi:2020phm,Bigazzi:2020avc,Ares:2020lbt}}. In contrast, the bubble wall velocity depends on out-of-equilibrium physics and its computation in terms of the microscopic theory is  challenging even for weakly coupled theories \cite{Moore:1995ua,Bodeker:2017cim,Hoeche:2020rsg,Vanvlasselaer:2020niz,Cai:2020djd}. The fact that the GW spectrum is particularly sensitive to this parameter \footnote{See e.g.~\cite{Hindmarsh:2016lnk,Hindmarsh:2019phv}} makes its computation not just challenging but pressing. 

The goal of this paper is to use holography to provide a first-principle calculation of the bubble wall velocity in a strongly-coupled theory with a gravity dual.\footnote{Holographic bubbles in the probe approximation have been considered in \cite{Li:2020ayr}.} We will not describe the nucleation of the bubble but focus on the post-nucleation dynamics and determine its dependence on the nucleation temperature. Holography maps the full, quantum-mechanical dynamics of the bubble to the classical dynamics of a black brane horizon in the dual geometry. We thus set up appropriate initial conditions for each nucleation temperature and numerically solve Einstein's equations to determine the time evolution of the bubble. From this we read off the velocity and the profile of the wall. We then verify that, in agreement with general expectations, ideal hydrodynamics describes the entire system except for the region near the wall.
\begin{figure}[t]
\center
\includegraphics[width=0.44\textwidth]{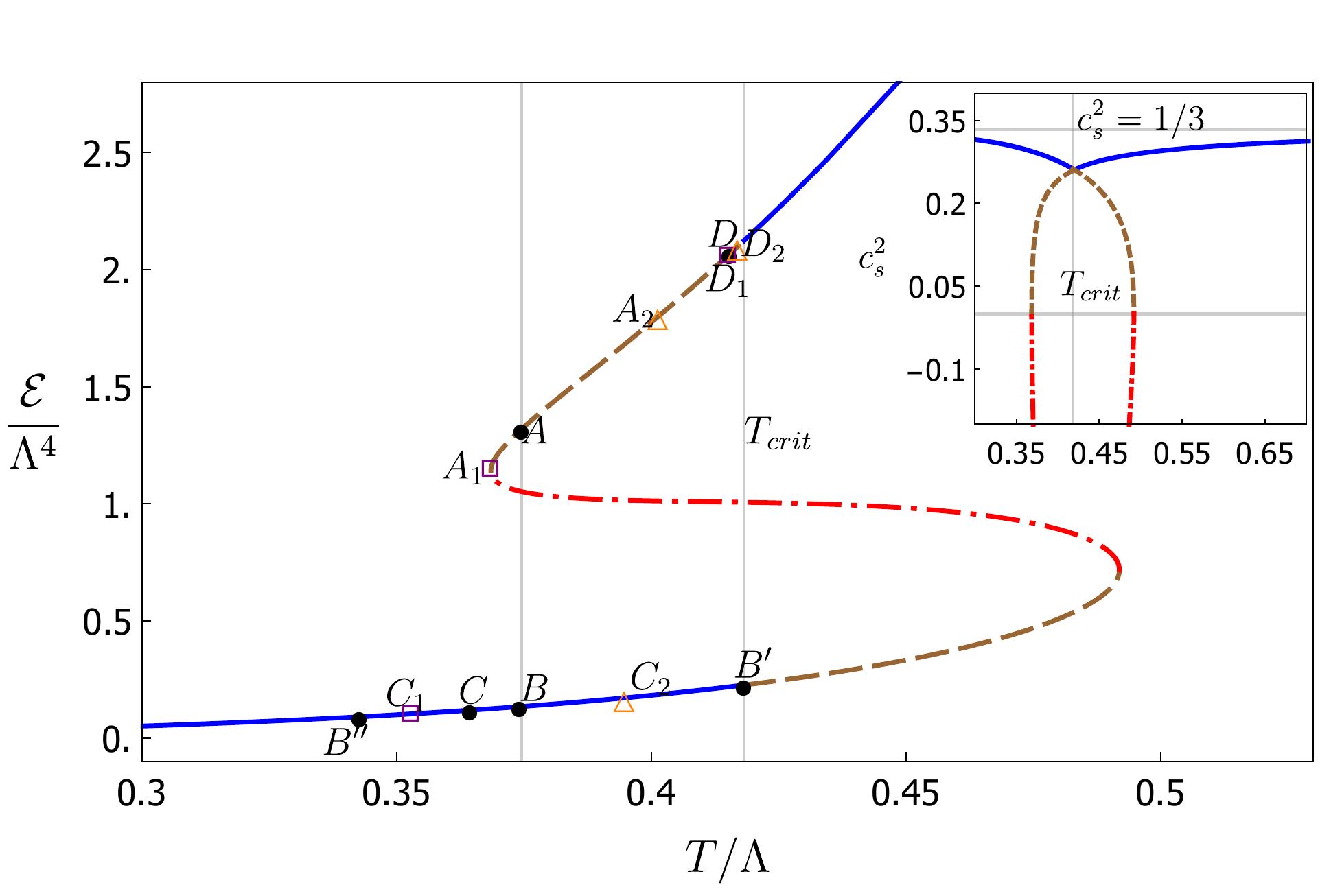}
\caption{\small\small Energy density as a function of temperature and (inset) speed of sound for $\phi_Q=10, \phi_M=0.85$. The grey vertical line on the right indicates the critical temperature at which the PT takes place. The grey vertical line on the left indicates that $A$ and $B$ have the same temperature. Stable states are shown in solid blue, metastable ones in dashed brown, and unstable ones in dotted-dashed red. 
}
\label{fig:Phase_Diagram}
\end{figure}

\noindent
{\bf Holographic model.}
We consider a five-dimensional Einstein-scalar model described by the action
\begin{equation}
S = \frac{2}{\kappa_5^2}\int d^5x\sqrt{-g}\left[\frac{1}{4}\mathcal{R}-\frac{1}{2}\left(\partial\phi\right)^2-V(\phi)\right]\,. 
\end{equation}
Exclusively for simplicity we assume that the scalar potential $V$ can be derived from a superpotential $W$ through the usual relation 
\be
\label{pot}
V(\phi)=-\frac{4}{3}W(\phi)^2+\frac{1}{2}W'(\phi)^2 \,.
\ee
Different choices of (super)potential correspond to different dual four-dimensional gauge theories. As in \cite{Bea:2018whf, Bea:2020ees}, we choose  
\begin{equation}
\label{superpot}
LW(\phi) = -\frac{3}{2}-\frac{\phi^2}{2}-\frac{\phi^4}{4\phi_M^2}+\frac{\phi^6}{\phi_Q},\\
\end{equation}
where $L$ is the asymptotic AdS radius and $\phi_Q$ and $\phi_M$ are constants. In the limit $\phi_Q\to \infty$ the sextic term is absent and the model reduces to that in \cite{Attems:2016ugt,Attems:2016tby,Attems:2017zam}. The motivation for the choice \eqref{superpot} is that this is a simple model of a non-conformal theory  with a first-order PT (for appropriate values of $\phi_Q$ and $\phi_M$ \cite{Bea:2018whf}) whose dual gravity solution is completely regular even at zero temperature.\footnote{In top-down deformations of $\mathcal{N}=4$ super Yang-Mills by relevant operators the field theory content includes both bosons and fermions. For example, the dimension-three operator dual to the bulk scalar field is typically a fermion bilinear. We expect the same to be generically true in bottom-up models like ours.} The fall-off of the scalar field near the  asymptotic AdS geometry determines the characteristic energy scale in the dual gauge theory, $\Lambda$, which in turn sets the value of the critical temperature, $T_c$. For concreteness, in  this paper we will focus on the model with  $\phi_Q=10, \phi_M=0.85$.\footnote{This choice is motivated by the requirement that the phase diagram be generic, meaning that it does not exhibit any large hierarchies, in contrast with e.g.~\cite{Attems:2017ezz,Attems:2019yqn}.}  The phase diagram for this case is shown in Fig.~\ref{fig:Phase_Diagram}, where we see the usual multi-valuedness characteristic of a first-order PT. The critical temperature is $T_c=0.418  \Lambda$, as indicated by the grey vertical line.

\noindent
{\bf Initial states.}
We now imagine that the system has been supercooled to some state $A$ on the upper metastable branch, and that at this point a bubble corresponding to some state $B$ on the lower stable branch is nucleated. The nucleation temperature is therefore $T_N=T_A$.  On general grounds we expect a non-zero probability to nucleate bubbles with different initial wall profiles and with different initial sizes. We will therefore vary these parameters and determine their effect on the subsequent post-nucleation dynamics. We will also vary the initial state  $B$ inside the bubble.  Although this is often assumed to have the same temperature as $A$, the initial-value problem with $T_B \neq T_A$ is perfectly well-defined on the gravity side. These parameters do not completely determine  the initial quantum state of the bubble. On the gauge theory side they only specify the one-point function of the stress-tensor  in the initial state, for example the profile of the energy density. On the gravity side they only specify the fall-off of the metric near the asymptotic AdS boundary. A complete determination of the initial quantum state requires knowledge of all the higher correlation functions in the gauge theory or, equivalently, the complete metric on the gravity side. Therefore we will also scan over different metrics in the initial data. As in \cite{Enqvist:1991xw,Ignatius:1993qn,Liu:1992tn}, for simplicity we will consider planar bubbles that are translationally invariant along the transverse $\{ x,y\}$-directions and expand only along the longitudinal $z$-direction. In particular, this means that there is no critical size for the bubble. We will report on spherical bubbles elsewhere \footnote{Y. Bea, J. Casalderrey-Solana, T. Giannakopoulos, D. Mateos, M. Sanchez-Garitaonandia, and M. Zilhao, work in progress}.

\noindent
{\bf Evolution.}
We begin with a bubble of size $14 \Lambda^{-1}$ and a state $B$ inside the bubble with temperature \mbox{$T_B=T_A=0.374\Lambda$}, as indicated in Fig.~\ref{fig:Phase_Diagram}. 
\begin{figure}[t]
\center
\includegraphics[width=0.42\textwidth]{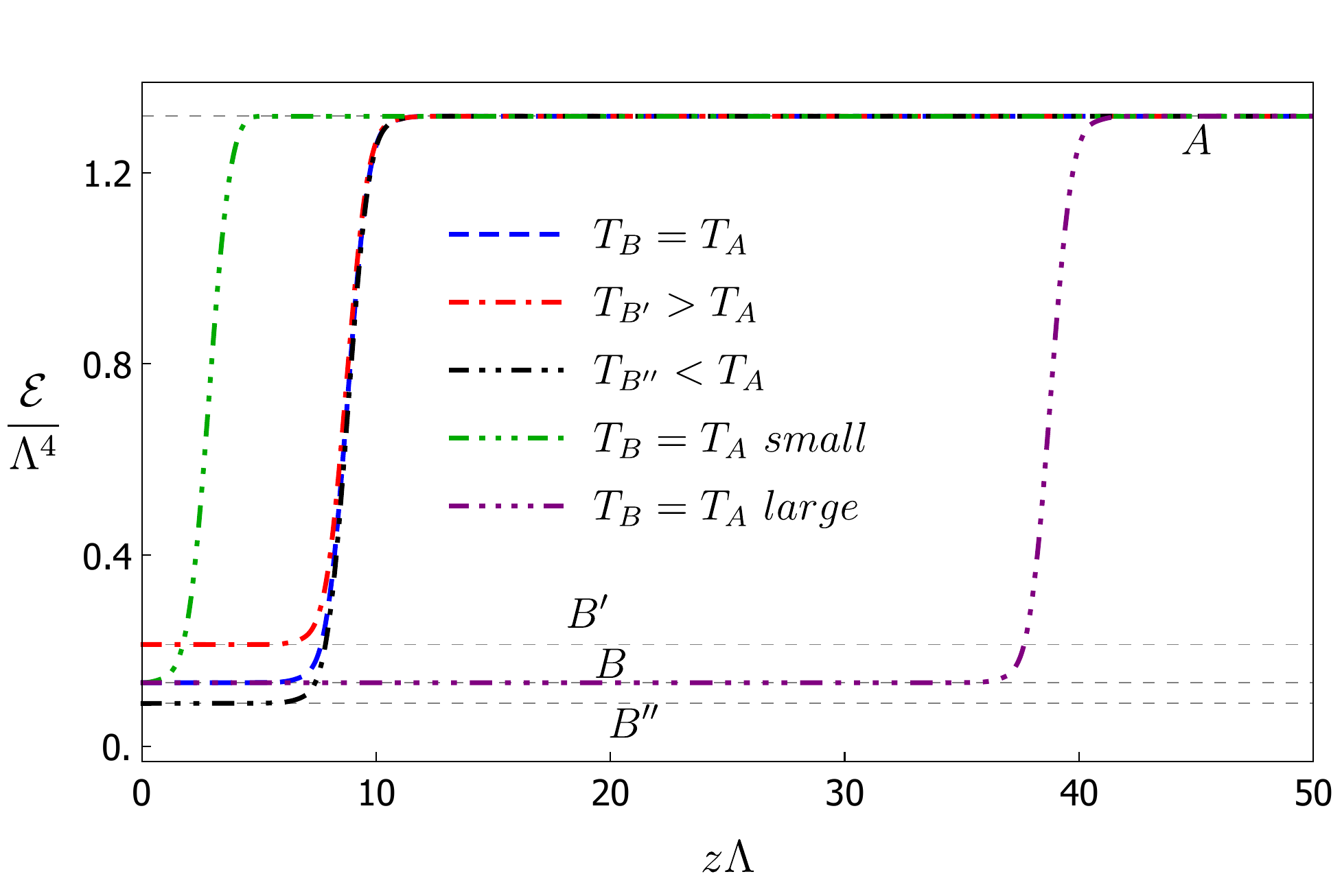}
\caption{\small\small Different initial energy profiles for the same nucleation temperature $T_A$. In this and in subsequent plots we only show positive values of $z$ because we only consider states invariant under $z \to -z$.}
\label{fig:initial_state}
\end{figure}
The initial energy profile, shown as a dashed blue curve in Fig.~\ref{fig:initial_state}, is arbitrarily chosen except for the fact  that it must interpolate between the energy $\mathcal{E}_B$ inside the bubble and the energy $\mathcal{E}_A$ outside the bubble.
\begin{figure}[t]
\center
\includegraphics[width=0.48\textwidth]{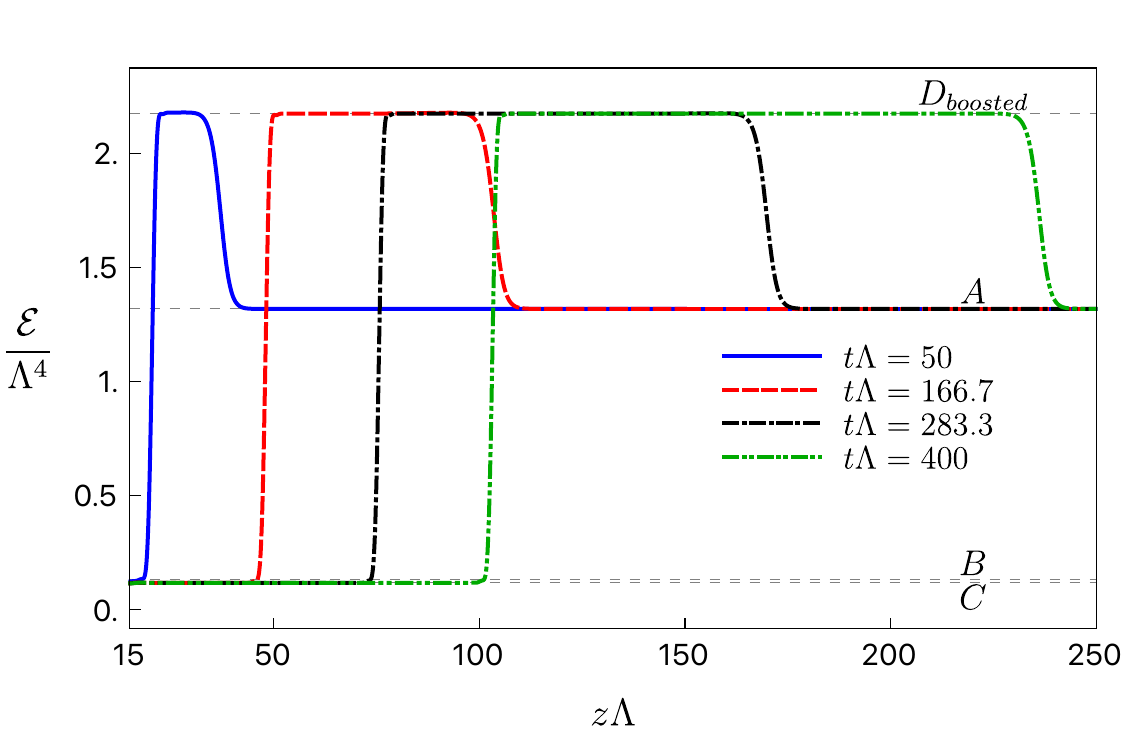}
\caption{\small\small Snapshots of the energy density profile at different times for the initial state with $T_B=T_A$ shown as a dashed blue curve in Fig.~\ref{fig:initial_state}. 
}
\label{fig:E_vs_z}
\end{figure}
We then solve the Einstein equations in the bulk along the lines of \cite{Chesler:2010bi,Chesler:2013lia,Attems:2017zam} to find the time evolution of this initial state. Since the pressure in $B$ is higher than in $A$, the initial wall is accelerated towards the right. Fig.~\ref{fig:E_vs_z} shows several snapshots of the resulting energy density at different times. 

\begin{figure}[t]
\center
\includegraphics[width=0.42\textwidth]{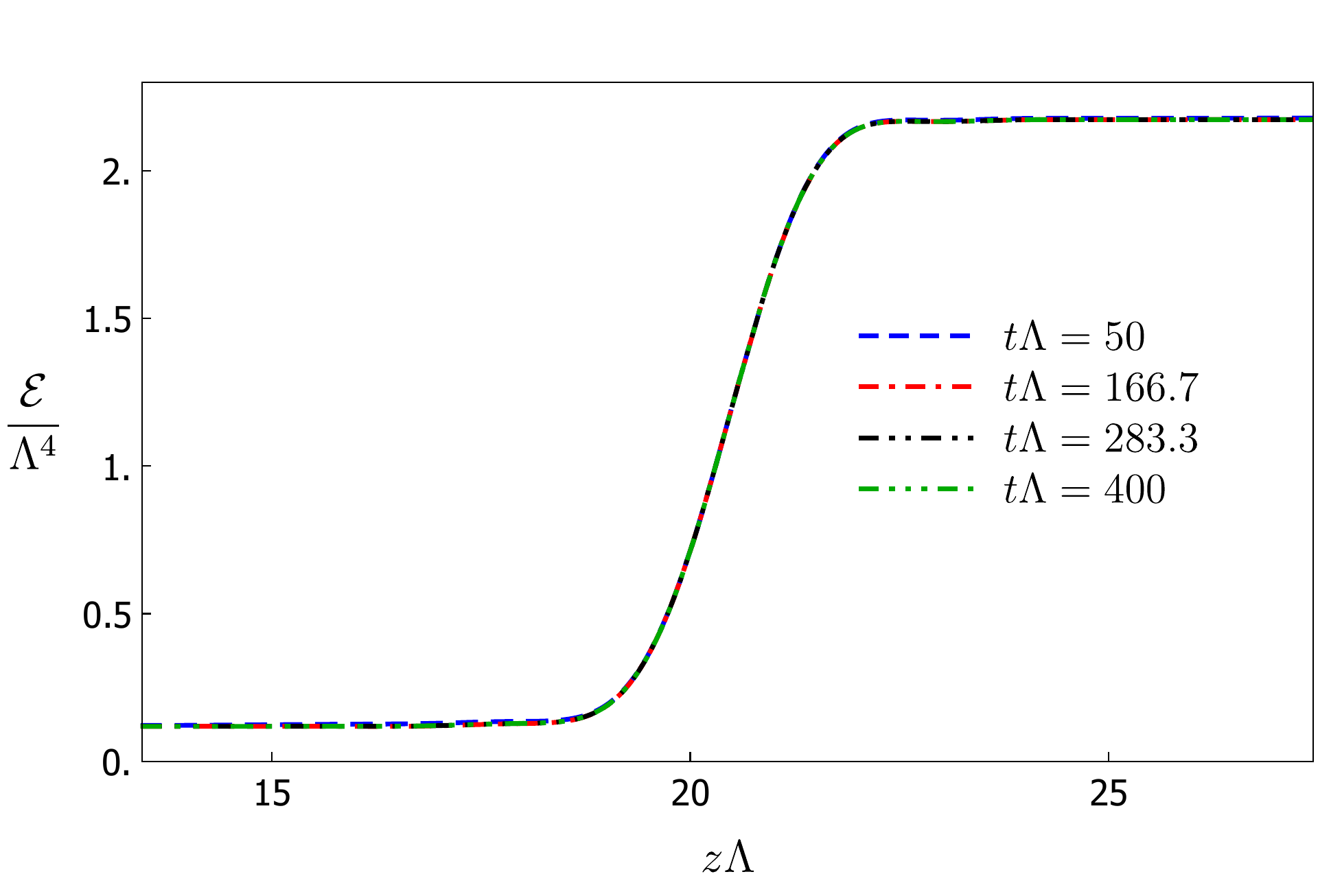}
\caption{\small\small Same wall profiles as in Fig.~\ref{fig:E_vs_z}, each shifted in $z$ by a different amount, to show that the wall profile remains constant in time.}
\label{fig:wall_shifted_1}
\end{figure}

\begin{figure}[t]
\center
\includegraphics[width=0.42\textwidth]{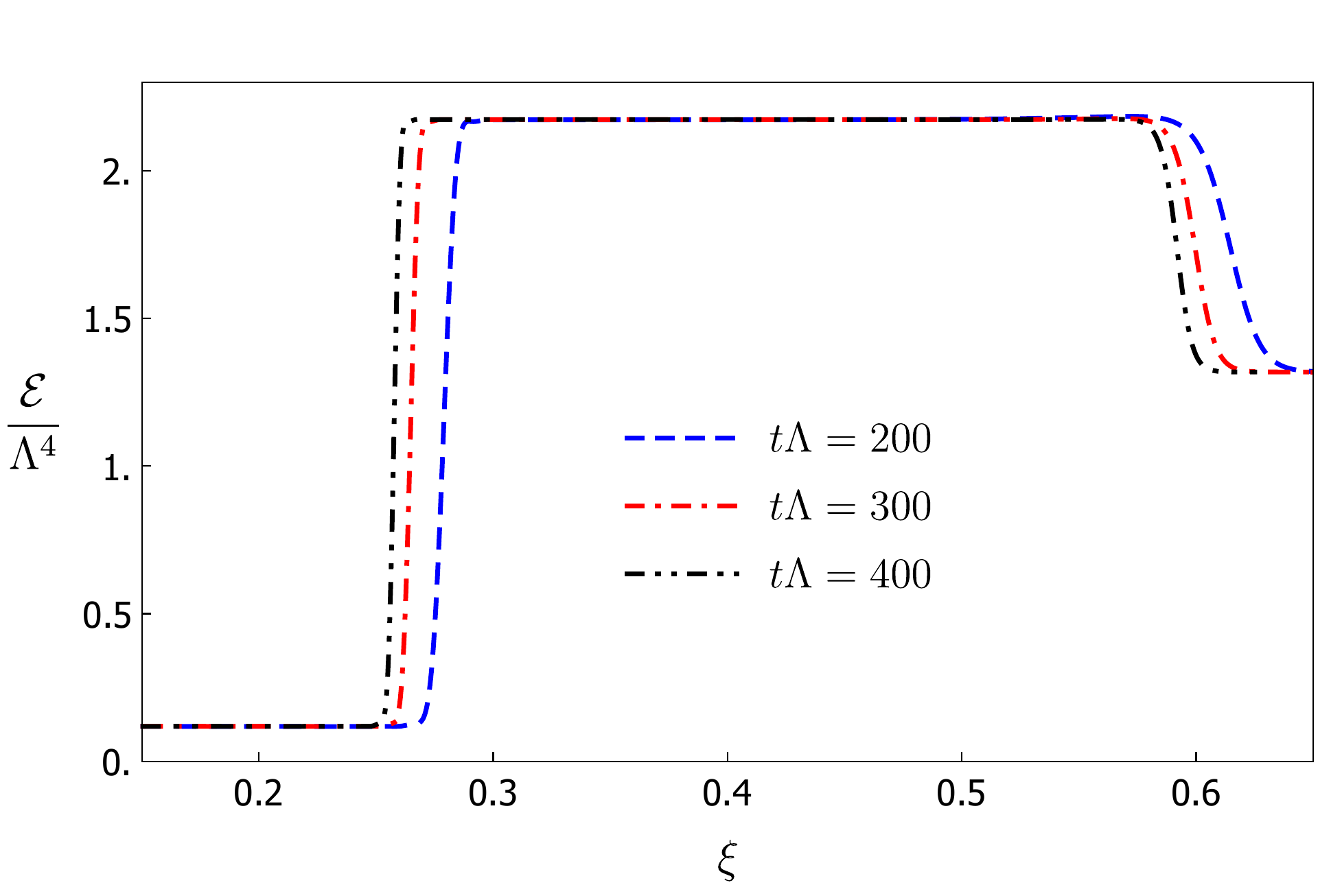}
\caption{\small\small Energy profile at as a function of $\xi = z/t$ for different values of $t$.}
\label{fig:selfsimilar}
\end{figure}

The wall starts at rest and reaches a steady state with terminal velocity  $v_{wall}^A \simeq 0.24$ in a time of order $ \Delta t \sim 50 \Lambda^{-1}$ \footnote{For holographic studies of non-equilibrium steady states in the context of the Riemann problem see 
e.g.~\cite{Ecker:2021ukv} and references thereof.}.  As illustrated by the solid blue curve in Fig.~\ref{fig:E_vs_z}, in this time the wall profile relaxes to a preferred shape. This shape remains constant at subsequent times, as shown by Fig.~\ref{fig:wall_shifted_1}. In addition, in this time the energy density inside the wall evolves to that of the state marked as $C$  in Fig.~\ref{fig:Phase_Diagram}. 
As time progresses, energy conservation implies that an intermediate ``hot'' region develops in between the wall and the asymptotic $A$-region. We have dubbed this region $D_{boosted}$ in Fig.~\ref{fig:E_vs_z}. The reason is that the fluid velocity in this intermediate region is constant and given by $v_D \simeq 0.22$, 
so $D_{boosted}$  is the state $D$ in Fig.~\ref{fig:Phase_Diagram} boosted to the right with velocity $v_D$.  The interface between the $D_{boosted}$- and the $A$-regions moves at constant velocity 
$v_{int} \simeq 0.57$. This means that the size of $D_{boosted}$  grows linearly with time. The width of the interface also grows, but more slowly than linearly. As a consequence, if we plot the energy profile in terms of $\xi = z/t$ for different fixed times, the interface between $D_{boosted}$ and $A$  approaches a discontinuity at late times, as illustrated in Fig.~\ref{fig:selfsimilar}. In this limit the  profile becomes a function of $\xi$ alone, as it is commonly assumed.  

\begin{figure}[t]
\center
\includegraphics[width=0.42\textwidth]{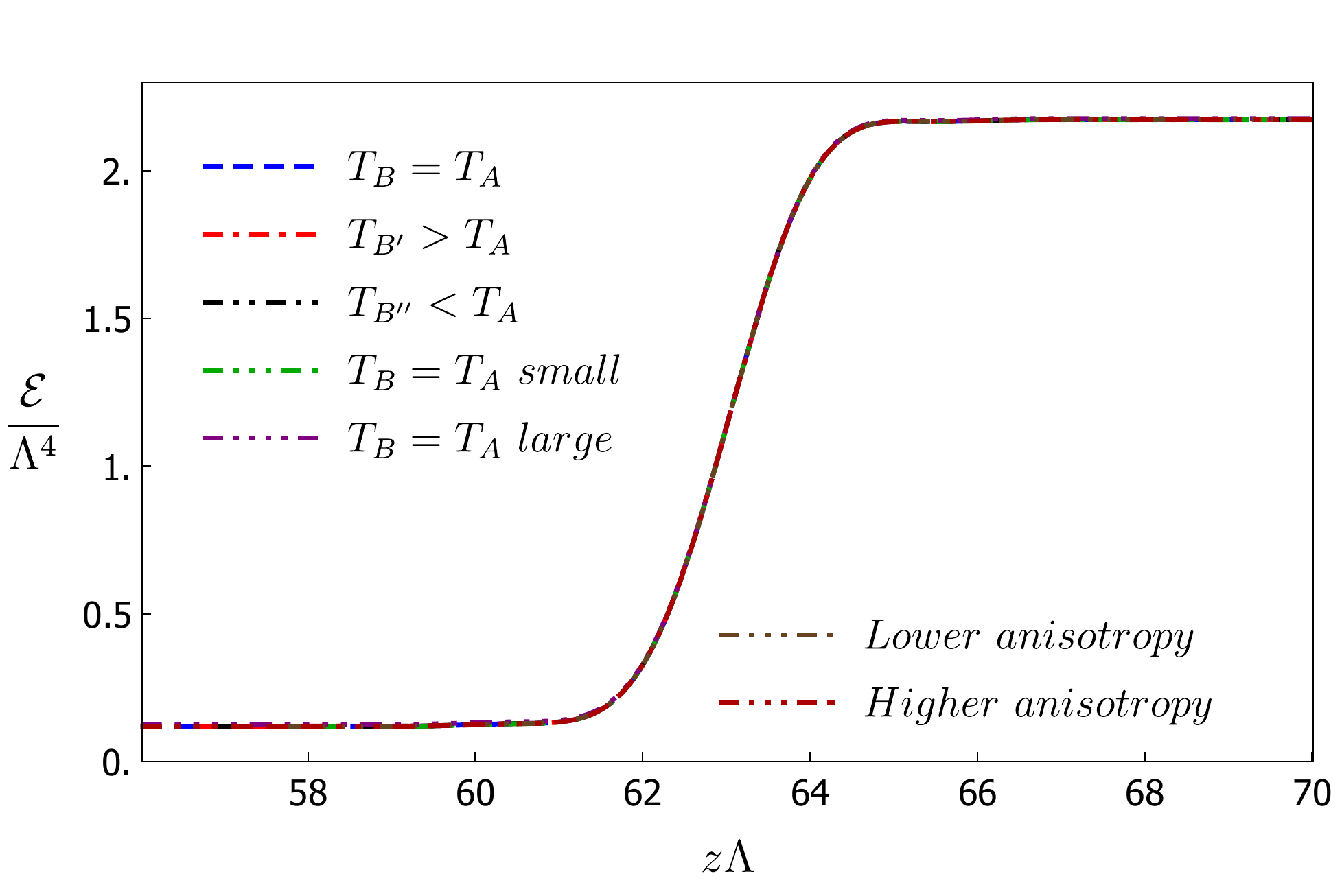}
\caption{\small\small Wall profiles for the same nucleation temperature $T_N=T_A$ but different initial conditions, each shifted in $z$ by a different amount to show that the wall profile is independent of the initial conditions. 
}
\label{fig:wall_diff_initial}
\end{figure}

The features of the late-time state such as the wall profile, the wall velocity, and the $C$- and $D$-states, are determined dynamically and are independent of the bubble initial conditions. We illustrate this for the wall profile in Fig.~\ref{fig:wall_diff_initial}. To obtain this plot we take the set of initial conditions above and vary one initial condition at a time to obtain a new set of wall profiles. Specifically, we change the initial state $B$ inside the bubble to the states $B'$ and $B''$ in Fig.~\ref{fig:Phase_Diagram}, whose corresponding energy profiles are shown in Fig.~\ref{fig:initial_state}. We also vary the initial size of the bubble to the larger and smaller values shown by the corresponding curves in Fig.~\ref{fig:initial_state}. In addition, the wall  profile for the smaller bubble is different from that of the original bubble. Finally, we change the initial bulk metric so as to increase or decrease the initial pressure anisotropy between the longitudinal and the transverse directions by an order of magnitude. As we see in Fig.~\ref{fig:wall_diff_initial}, all these changes result in the same late-time wall profile. 

We now turn to the dependence on the nucleation temperature. The states $C$ and $D$ and the wall velocity vary monotonically with $T_N$. Indeed, as $T_N$ approaches $T_{crit}$ from the left  the states $C$ and $D$ approach the vertical line at $T=T_{crit}$, the wall velocity goes to zero and the system approaches a static, phase-separated configuration in which the states inside and outside the bubble coexist at the critical temperature \cite{Attems:2019yqn}. In the opposite limit, as $T_N$ decreases from $T_{crit}$ towards the end of the metastable branch, labelled as $A_1$ in Fig.~\ref{fig:Phase_Diagram}, the states $C$ and $D$ move to the left and approach $C_1$  and $D_1$, respectively. Similarly,   the wall velocity increases monotonically from zero to a maximum value $v_{wall}^{A_1} \simeq 0.29$. 

\begin{figure}[t]
\center
\includegraphics[width=0.40\textwidth]{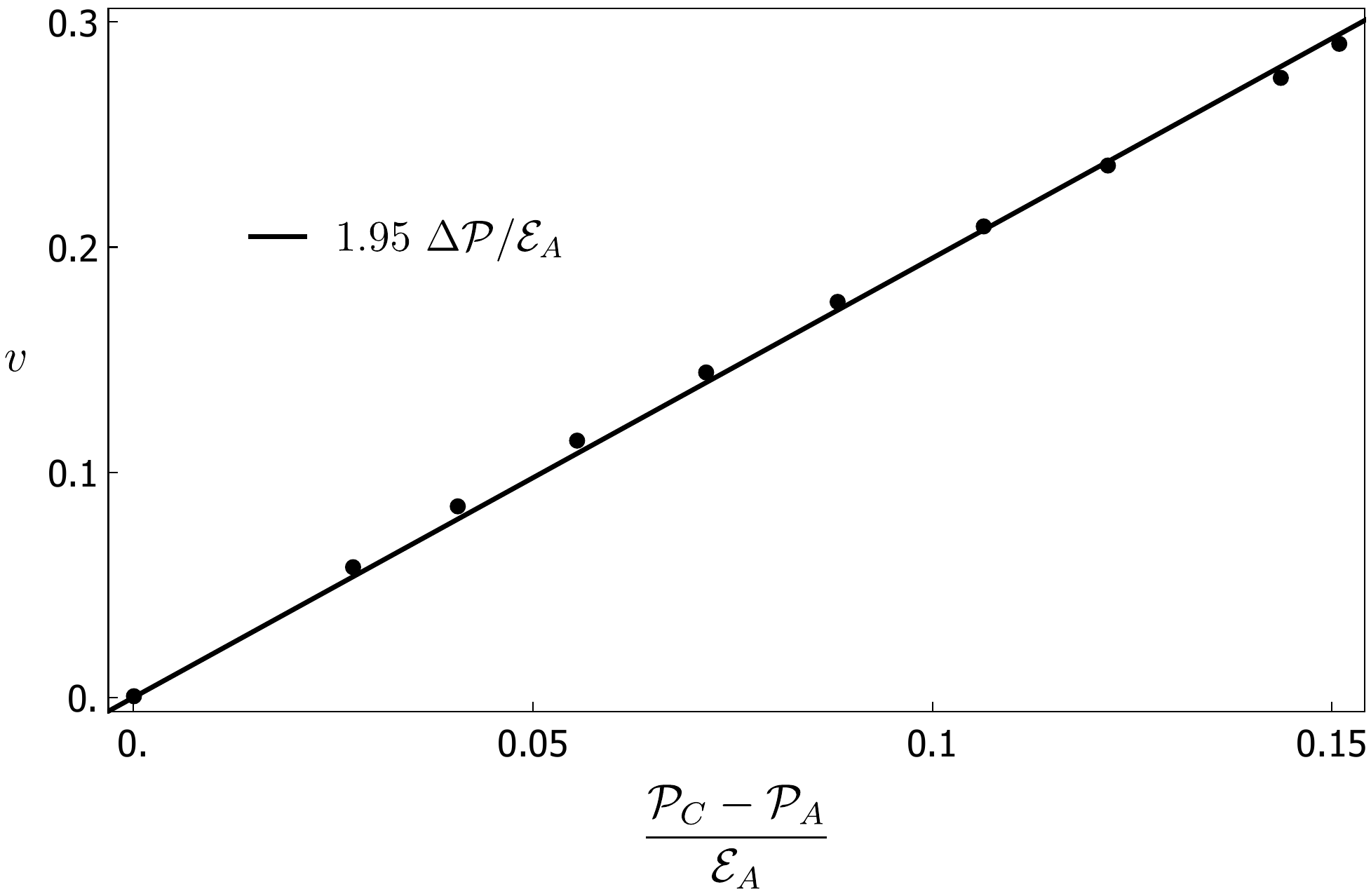}
\caption{\small\small The points show the wall velocity for different nucleation temperatures. The line is a  fit as a function of the ratio between the pressure difference inside and outside the bubble and the energy density outside the bubble.  
}
\label{fig:linear_v_wall}
\end{figure}

We have explored the dependence of $v_{wall}$ on different properties of the state $A$. The most suggestive result is shown in Fig.~\ref{fig:linear_v_wall}, which seems to imply a linear dependence on the ratio between the pressure difference inside and outside the bubble and the energy density outside the bubble. Heuristically, this relation seems plausible given that the force trying to accelerate the bubble increases with the pressure difference, whereas the resisting force  grows with the energy density outside the bubble. 

\begin{figure}[t]
\center
\includegraphics[width=0.42\textwidth]{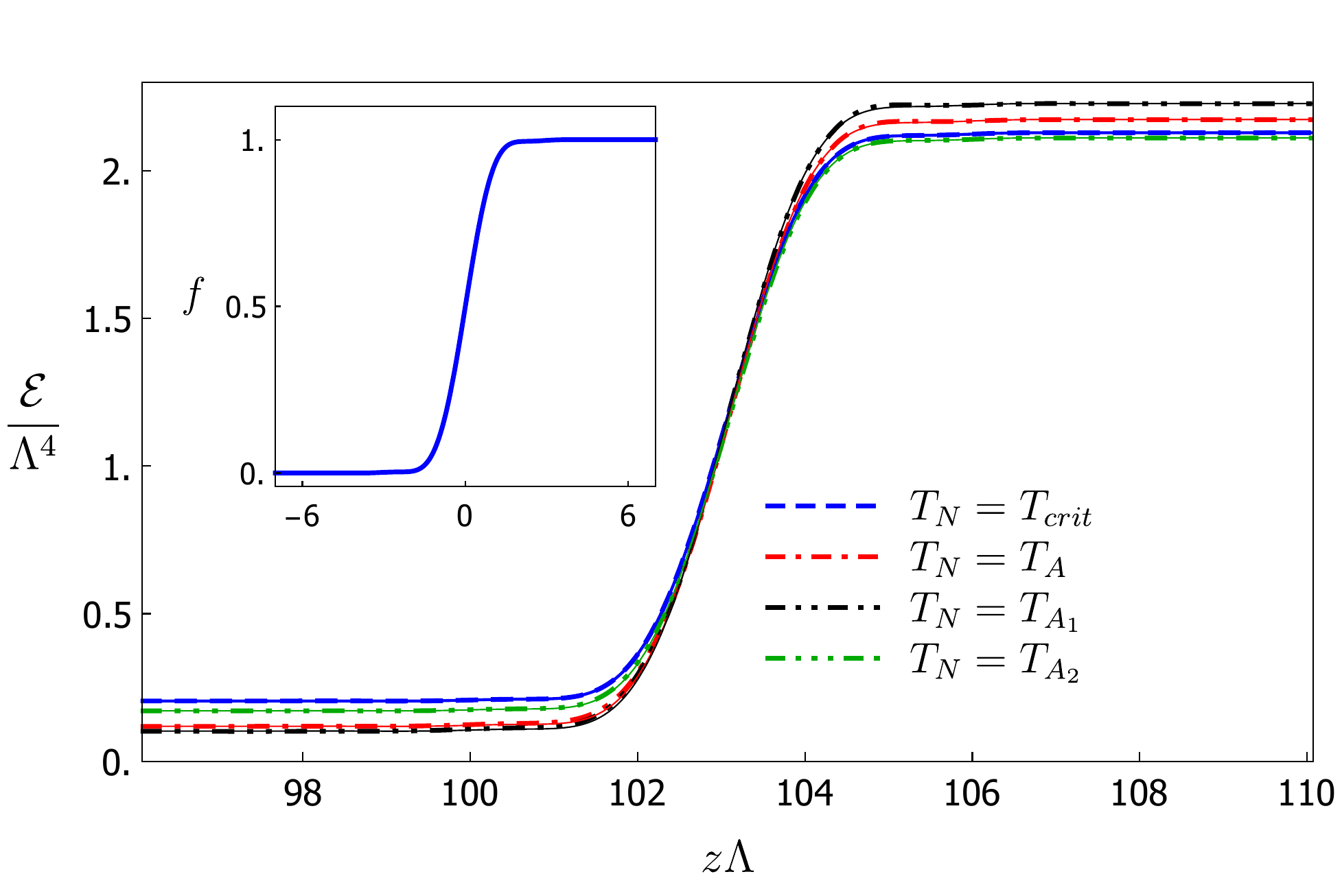}
\caption{\small\small Comparison between the exact wall profiles for several nucleation temperatures with the results of applying Eq.~(\ref{ez}). The case $T_N=T_{crit}$ corresponds to a static, phase-separated configuration. The universal function $f$ is shown in the inset. 
}
\label{fig:stationary_scaled}
\end{figure}

Changing the nucleation temperature also changes the wall profile. However, we empirically observe that, up to a rescaling, the latter is well approximated by the interface of a phase-separated configuration at $T=T_{crit}$ \cite{Attems:2019yqn}. Specifically, the wall  profile for any $T_N$ is given by 
\be
\label{ez}
\mathcal{E}(z) = \mathcal{E}_C + (\mathcal{E}_D-\mathcal{E}_C) f(\Lambda z) \,,
\ee
where the energies of the $C$- and $D$-states depend on $T_N$ but $f$ is a $T_N$-independent, universal function that only depends on the theory. In particular, taking $T_C=T_D=T_{crit}$, this formula gives the profile of the phase-separated configuration. The latter is shown in Fig.~\ref{fig:stationary_scaled}, where we also compare the exact wall profiles for several nucleation temperatures with those predicted by Eq.~(\ref{ez}).

\begin{figure}[t]
\center
\includegraphics[width=0.42\textwidth]{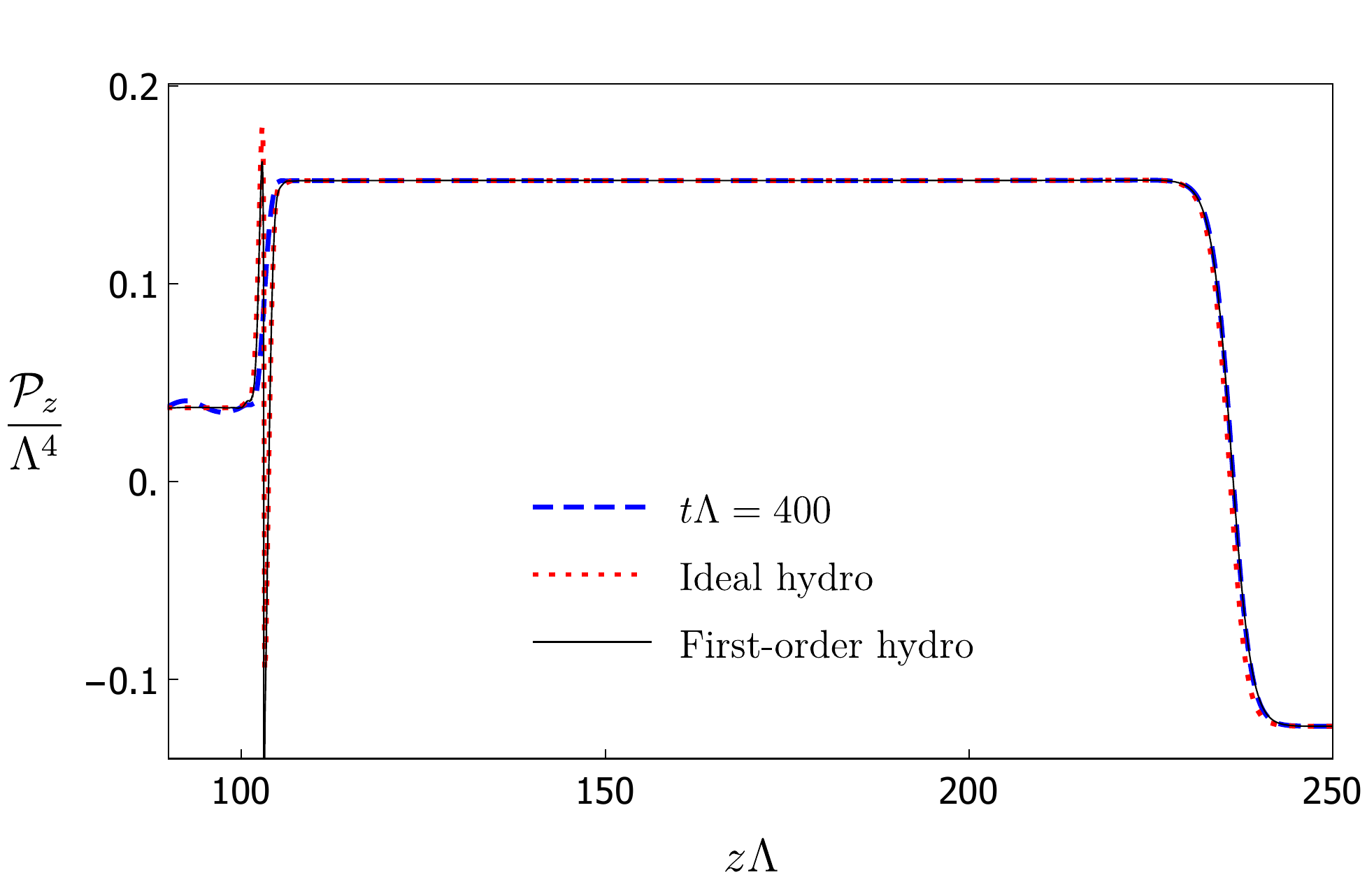}
\caption{\small\small Comparison between the holographic longitudinal pressure at late times and the ideal and the first-order hydrodynamic predictions.}
\label{fig:hydro}
\end{figure}

\noindent
{\bf Hydrodynamics.}
As the bubble expands the gradients away from the wall get diluted. Therefore the late-time state is expected to be well described by ideal hydrodynamics everywhere except in the region near the wall. This is confirmed by Fig.~\ref{fig:hydro}, where we compare the exact result for the longitudinal pressure with the prediction of both ideal and first-order viscous hydrodynamics at late times. We see that none of the hydrodynamic curves describe the wall region correctly. 
Nevertheless, at asymptotically late times the size of the wall becomes negligible and we can use ideal hydrodynamics to constraint the properties of the bubble. At those  times  we can treat both the wall and the interface between the $D_{boosted}$- and 
$A$-regions as discontinuities and assume that the physics only depends on $\xi$. Requiring that the energy and momentum fluxes are the same on both sides of these discontinuities leads to a set of matching conditions (see e.g.~\cite{Espinosa:2010hh}). In combination with the hydrodynamic equations away from the wall, these conditions determine the $C$- and 
$D$-states, the fluid velocity in $D$, and the velocity of the interface in terms of the nucleation temperature and the wall velocity.  This means that, for a given nucleation temperature, the entire system is controlled by the wall velocity.

\noindent
{\bf Discussion.}
We have used holography to determine the post-nucleation dynamics of bubbles in a theory with a first-order PT. The state $C$ inside the bubble, the state $D$ between the bubble wall and the asymptotic region, the wall velocity and the wall profile are all dynamically determined and are independent of the initial conditions. In general, the temperature in  $C$ differs  from the nucleation temperature. 

One interesting feature of our  model is that the speed of sound is not constant. When the nucleation temperature approaches the critical temperature the wall velocity approaches zero. As a consequence the bubble expansion is guaranteed to be a deflagration. In the opposite limit, as $A$ approaches the turning point labelled $A_1$ in Fig.~\ref{fig:Phase_Diagram}, the speed of sound in $A$ goes to zero. This guarantees that $v_{wall} > c_s^A$. Nevertheless, in our model  the wall velocity remains lower than the speed of sound in $D$, since $v_{wall}^{A_1} \simeq 0.24$ and $c_s^{D_1} \simeq 0.51$, and we still find a deflagration with zero fluid velocity behind the wall. 

It will be interesting to construct holographic models in which the wall velocity is much larger that the speed of sound in any of the regions, as expected in a detonation. On the one hand, the  approximate linear relation that we uncovered in Fig.~\ref{fig:linear_v_wall} suggests that this will require that the pressure difference between the inside and the outside of the bubble be comparable to the energy density outside the bubble. On the other hand we emphasize that, at this point, this linear relation should be taken purely as an empirical observation in a single model. Establishing its validity beyond this case requires further analysis. 

Although the $C$- and $D$-states depend on the nucleation temperature we found evidence that, up to a rescaling, the wall profile is well approximated by a universal, $T_N$-independent function, as illustrated in Fig.~\ref{fig:stationary_scaled}. However, the small deviations from this universal form grow with the wall velocity. This suggests that the good agreement seen in Fig.~\ref{fig:stationary_scaled}  may be due to the relatively low velocity of the wall in our model, and that this agreement may or may not persist in cases with higher velocities.   

As expected on general grounds, we verified that the entire system except for the region near the wall is well described by ideal hydrodynamics at late times. This, together with the matching conditions across the wall, determines the properties of the entire system in terms of the  nucleation temperature and the wall velocity. In order to further determine the velocity in terms of the nucleation temperature a model capable of resolving the dynamics in the wall region is needed. Here we have used the microscopic description provided by holography. The fact that the wall profile is related  to the interface of a phase-separated configuration \cite{Attems:2019yqn} suggests that an effective description of the wall dynamics, based on the ``purely spatial'' formulation of second-order hydrodynamics, may also be possible. 

\noindent
{\bf Acknowledgements.}
We are grateful to Alessio Caddeo, Mark Hindmarsh, Martin Sasieta and Marija Toma\v{s}evi\'{c} for discussions. 
YB is supported by the European Research Council Grant No.~ERC-2014-StG 639022-NewNGR. TG acknowledges financial support from FCT/Portugal Grant No.~PD/BD/135425/2017 in the framework of the Doctoral Programme IDPASC-Portugal.
MSG acknowledges financial support from the APIF program, fellowship APIF\_18\_19/226. 
MZ acknowledges financial support provided by FCT/Portugal through the IF
programme, grant IF/00729/2015.  JCS, DM and MSG are also supported by grants SGR-2017-754, CEX2019-000918-M, PID2019-105614GB-C21 and PID2019-105614GB-C22. The authors thankfully acknowledge the computer resources, technical expertise and assistance provided by CENTRA/IST. Computations were performed in part at the cluster ``Baltasar-Sete-S\'ois" and supported by the H2020 ERC Consolidator Grant ``Matter and strong field gravity: New frontiers in Einstein's theory" grant agreement No.~MaGRaTh-646597. We also thank the MareNostrum supercomputer at the BSC (activity Id FI-2020-1-0007) for significant computational resources.

\bibliographystyle{apsrev4-1}
\bibliography{refs_Bubble_Wall_Velocity_from_Holography}

\end{document}